\documentclass[aps,prb,twocolumn]{revtex4}
\usepackage{epsfig}

\begin{document}

\title{Proper weak-coupling approach to the 
       periodic s-d(f) exchange model}
\author{T.~Hickel} 
\email{hickel@physik.hu-berlin.de}
\author{W.~Nolting}
\affiliation{
    Festk\"orpertheorie, Institut f\"ur Physik,
    Humboldt-Universit\"at, 12489 Berlin, Germany}
\date{\today}

\begin{abstract}
 The periodic s-d(f) exchange model is characterized by a wide variety
 of interesting applications, a simple mathematical structure
 and a limited number of reliable approximations which take 
 care of the quantum nature of the participating spins. We suggest 
 the use of a projection-operator method for getting information 
 perturbationally, which are not accessible via diagrammatic
 approaches. In this paper we present in particular results beyond 
 perturbation theory, which are obtained such that almost all exactly 
 known limiting cases are incorporated correctly. 
 We discuss a variety of possible methods and evaluate their
 consequences for one-particle properties. These
 considerations serve as a guideline for a more effective approach 
 to the model. 
\end{abstract}

\newcommand{\Ham}{{\cal H}}
\newcommand{\CF}[1]{\left\langle#1\right\rangle}
\newcommand{\n}{{\hat n}^{ } }
\newcommand{\cd}[1]{c^\dagger_{#1}}
\newcommand{\cn}[1]{c^{ }_{#1}}
\newcommand{\dn}{\downarrow}
\newcommand{\up}{\uparrow}
\newcommand{\ps}{\sigma}
\newcommand{\ms}{{-\sigma}}
\newcommand{\ie}{{\rm i}}
\newcommand{\Jt}{\tilde J}
\newcommand{\eps}{\varepsilon}
\newcommand{\one}{1\hspace{-0.3em}{\rm l}}

\newcommand{\LSP}[2]{\left(#1\left|#2\right.\right)}
\newcommand{\Lket}[1]{\left|#1\right)}
\newcommand{\Lbra}[1]{\left(#1\right|}
\newcommand{\ket}[1]{\mbox{$\, \mid \! #1 \rangle$}}
\newcommand{\Lio}{{\cal L}}
\newcommand{\bk}{{\bf k}}
\newcommand{\bp}{{\bf p}}
\newcommand{\bq}{{\bf q}}
\newcommand{\bz}{{\bf 0}}

\maketitle
\section{Intention}

 The interplay of localized magnetic moments (of d- or f-type) with 
 itinerant s-electrons of a partially filled conduction band is of 
 indisputable
 importance for the explanation of many effects in condensed matter 
 physics \cite{Zen51,AH55}. In recent years and in the context of 
 diluted magnetic semiconductors it even became a driving mechanism 
 for electronic applications (spintronics) \cite{Aka98,DOM00}.
 For a theoretical description the Hamiltonian
 \begin{equation}
   \label{eq:Ham1}
   \Ham = \Ham_0 + \Ham_{\rm sf}
        = \sum_{i,j} \sum_\ps T_{ij} \, \cd{i\ps} \cn{j\ps} 
        - J \sum_i \mbox{\boldmath$\sigma$\unboldmath}_i \!\cdot\! {\bf S}_i .
 \end{equation}
 is probably the simplest choice possible. The localized magnetic moments
 are represented by quantum-mechanical spin operators ${\bf S}_i$ and 
 interact with the spin of the conduction electrons 
 $\mbox{\boldmath$\sigma$\unboldmath}_i$ via an exchange interaction
 $J$. The dispersion $\eps_\bk$ of the latter is determined by 
 the hopping integrals $T_{ij}$. 
 In recent years one has referred to the Hamiltonian (\ref{eq:Ham1})
 as the Kondo-lattice model, nevertheless sf-, sd- or
 (for very large $J$) double-exchange model are probably more suitable
 descriptions. 

 Despite the large variety of systems to which the s-f model can be applied 
 \cite{NRRM01}, a convincing approximation scheme is still missing. 
 Previous attempts developed along two principle directions: On the one hand 
 the Green's function hierarchy of equations of motions has been 
 restricted by some decoupling schemes \cite{NRMJ97}. 
 Even though these efforts allowed a fair description of magnetic
 semiconductors such as europium chalcogenides \cite{MN02}, this kind of 
 approximation always suffers from a lack of controllability. 
 On the other hand, a classical treatment of the localized spins 
 lead to some substantial results obtained using
 dynamical mean field theory \cite{HV00} or Monte Carlo techniques 
 \cite{KPEL02}. However, it has been shown that the quantum-mechanical
 character of the spins has indeed a substantial impact on the electronic 
 properties of the concerned materials \cite{Edw02,MSN01}. 

 Therefore, an analytical and numerical treatment of the s-f model should be
 aspired to, which retains the quantum nature of the spins and ensures the
 controllability. A perturbational approach would be a good candidate
 for the latter. However, the incorporation of quantum-spin operators 
 causes difficulties since Wick's theorem, which is generally used to
 evaluate Feynman diagrams, is not applicable to this class of
 operators. This is probably the reason why up to now only 
 few attempts exist in this direction. These are mostly limited to special
 situations like one-dimensional systems or half-filling \cite{ShW99,Gu02}. 

 We suggest to circumvent these difficulties by using the 
 projection-operator method (POM) as introduced by Mori
 \cite{Mor65a,Mor65b}.  Its goal is the expansion of resolvents 
 such as the one-particle Green's functions in a continued fraction. 
 The further one goes in the expansion of the continued fraction, 
 the higher the accuracy of the results obtained. It has already 
 been applied successfully for a 
 weak-coupling approach to the Hubbard model where it leads to 
 convincing results \cite{Kis87,BuJ90}. Its application to the 
 s-f model would be an obvious development, even though straight 
 forward test calculations were not very successful.\cite{MSN01}

 A weak-coupling approach to the s-f model is undoubtedly very
 interesting, both from an experimental and a theoretical point of 
 view. First of all, the parameter regime is probably important 
 for diluted magnetic semiconductors. Secondly, 
 the well-known RKKY exchange mechanism \cite{RuK54,Kas56,Yos57}, 
 usually used for a qualitative description of magnetism in these 
 materials, is also nothing else but the application of second order 
 perturbation theory to Hamiltonian (\ref{eq:Ham1}). However, the goal
 of this approach is the determination of the ground state energy, 
 leading to an effective Heisenberg interaction. In contrast to that,
 here we will show that we are able to analyze the much richer spectrum
 of dynamical properties of the s-f model to the same order of the coupling 
 constant and beyond it. 

 Whenever a second order perturbation theory (SOPT) description is performed
 in many-body theory there are principally three different ways
 of treating emerging propagators: a {\it conventional} SOPT uses only 
 free propagators, a SOPT {\it relative to Hartree-Fock} (HF) replaces these
 propagators by the corresponding mean-field expressions and a 
 {\it self-consistent treatment} of the SOPT replaces all propagators by 
 (functionals of) the full propagators as obtained in
 the previous step of iteration. A priori it is usually not known 
 which version yields the most reliable results. 
 Of course, the {\it self-consistent} version includes a summation of 
 more diagrams than the other methods. However, since only a partial class of 
 diagrams is summed, it is unclear which important diagrams are being
 missed out or canceled, double-counted or even taken wrongly.
 There are more profound considerations as put forward by Kadanoff and
 Baym \cite{BK61} arguing that the self-consistent approach is a
 conserving approximation which automatically satisfies the Luttinger
 theorem \cite{LW60} and Fermi-liquid properties. However, it is uncertain
 to which degree this applies to a model which is not exclusively
 composed of fermions. 

 Even for more established models such as the periodic Anderson model (PAM), 
 the Falicov-Kimball model (FKM) or the Hubbard model (HM) the question
 of the most appropriate version has been discussed intensively. 
 For the PAM Yamada and Yosida \cite{YY70,YY75a,YY75b} started the 
 perturbational investigations directly by considering deviations 
 from the non-magnetic HF solution. Later Schweitzer 
 and Czycholl \cite{SC90} were able to numerically compare this 
 approach with a self-consistent SOPT. Even though this version obeys
 more of the Luttinger sum rules \cite{LW60}, the self-consistent 
 version (in contrast to the version relative to HF) 
 failed to show the one-particle peaks near $E_f$ and $E_f+U$
 in the $f$-electron spectral function. The FKM is another example
 of a model where a not fully self-consistent SOPT treatment 
 qualitatively reproduces exact results, whereas the self-consistent
 SOPT does not. \cite{SC90} 
 For the HM the situation is slightly more complicated. One 
 can show that SOPT relative to HF does not yield a metal-insulator 
 transition and does not show a breakdown of the Fermi-liquid 
 behavior.\cite{WeC94} On the other hand, a straightforward
 application of a self-consistent SOPT does not reproduce 
 the Hubbard bands in the atomic limit. \cite{MH89} More
 sophisticated methods such as the interpolation scheme
 of Edwards and Hertz \cite{EdH90,WeC94} (a version relative to
 HF) or the iterative perturbation theory (IPT) of Georges 
 and Kotliar \cite{GeK92} (a self-consistent version) are required. 
 For the latter approach the restriction to half-filling has been
 removed by the modified perturbation theory (MPT) of Kajueter,
 Kotliar \cite{KK96} and Potthoff, Wegner, Nolting 
 \cite{PWN97}. The MPT is probably the most convincing
 analytical approach to the HM.\cite{PHWN98}

 With the present paper we extend this kind of discussion to the
 electronic part of the periodic s-d(f) exchange model. We will argue that
 it is indeed a {\it self-consistent} ansatz for the electronic
 self-energy which is the most promising for this model. Other
 possible weak-coupling approaches are ruled out after a 
 direct comparison with the results of our method of choice. 
 It can be shown that minor changes in the analytical 
 method have drastic effects on one-particle properties,
 such as the density of states. We believe that a more profound analysis
 of the s-f model (e.g. by a combination with band-structure calculations)
 can be based on these considerations.

 The article is organized as follows: As a starting point we 
 derive  in section \ref{sc:method} an SOPT for the s-f model which 
 makes use of the POM. In section \ref{sc:exact} we will study
 in some detail the exactly soluble limit of a single conduction 
 electron in a ferromagnetically saturated semiconductor.
 This limit is an excellent testing ground for the 
 implementation of the POM. Even more important, the
 experience with other models shows that it is indispensable
 to have non-perturbative, exact statements which can be used 
 to judge the quality of the results obtained. In a next step 
 (section \ref{sc:arbitrary}) the experiences for 
 an improvement of this limit are generalized to arbitrary parameter
 configurations. With all these preparations we are able to
 present the results of the self-consistent calculation and 
 a comparison with other methods in sections \ref{sc:numerics}
 and \ref{sc:correct}. The findings are summarized in section
 \ref{sc:summary}. 

 It is worth mentioning, that our approach seems to be
 related to a moment-conserving interpolation scheme of the self-energy
 as published by Nolting {\it et al.} \cite{NRRM01,NRR03}.
 There, a general structure of the electronic self-energy, which looks
 similar to the one presented here, has been found by systematically 
 studying all known exact statements on the s-f model. However, their 
 analysis is focused on the low-density limit and ensures the
 correctness of these statements for $n \to 0$ (or $n \to 2$) only. 
 In contrast to that the approach given in this paper concentrates on 
 the weak-coupling behavior. Indeed, independent of
 the occupation number the correctness of the self-energy up to order 
 $\Jt^2$ in the coupling parameter and up to order $E^{-2}$ in the
 high-energy expansion is guaranteed! Additionally, we can fulfill the same
 criteria for $n \to 0$ as given in the above mentioned
 publication. Nevertheless, the two approaches are not identical
 even for $n=0$, but otherwise arbitrary parameters.

\section{Second order perturbation theory}
\label{sc:method}

 As mentioned above, we use the projection-operator method
 (POM) \cite{Mor65a,Mor65b} since it allows an expansion of resolvents 
 without the use of Wick's theorem. The approximation consists in
 considering only a physically relevant subspace of the Liouville
 space. With the simplest choice the Liouville space is spanned by 
 single-particle states $\mid\! \cd{\bk\ps} )$. Accordingly the projection
 operator and its orthogonal complement are defined as
 \begin{equation}
   \label{eq:POM-P&Q}
   P = \Lket{\cd{\bk\ps}}\Lbra{\cd{\bk\ps}} \quad \mbox{and} \quad
   Q = 1 - \Lket{\cd{\bk\ps}}\Lbra{\cd{\bk\ps}}.
 \end{equation}
 These definitions require the existence of a scalar product, which in
 our calculations is conveniently chosen to be the  
 thermodynamic average:
 \begin{equation}
   \label{eq:POM-LSP}
   \LSP{A}{B} \equiv \CF{\left[A^+;B\right]_+}. 
 \end{equation}

 Within the POM the one-particle Green's function is given by
 the following dynamical equation
 \begin{equation}
   \label{eq:POM-GF}
   G_{\bk\ps} 
      = \Lbra{\cd{\bk\ps}} \frac{1}{\omega-\Lio} \Lket{\cd{\bk\ps}}
      = \frac{\chi_{\bk\ps}}{\omega - [\Omega_{\bk\ps} 
              + M_{\bk\ps}(\omega)] \chi_{\bk\ps}^{-1} } ,
 \end{equation}
 where $\omega= E + \ie 0^+$ and the Liouville operator $\Lio$ 
 with its property 
 \( \Lio \Lket{A} \equiv \Lket{[ \Ham, A ]_-} \)
 has been incorporated. For the choice (\ref{eq:POM-P&Q}) the 
 susceptibility matrix $\chi_{\bk\ps}=(\cd{\bk\ps} | \cd{\bk\ps} )$
 is particularly simple: $\chi_{\bk\ps} \equiv 1$.
 The frequency matrix
 \begin{equation}
   \label{eq:POM-frequency}
   \Omega_{\bk\ps} = \Lbra{\cd{\bk\ps}} \Lio \Lket{\cd{\bk\ps}}
         = \eps_\bk - \Jt \, z_\ps\!\CF{S^z},
 \end{equation}
 on the other hand, corresponds to the mean-field result 
 \begin{equation}
   \label{eq:MF-result}
   G_{\bk\ps}^{\rm (MF)} (\omega) 
   = \frac{1}{\omega - \eps_\bk + \Jt \,z_\ps\!\CF{ S^z } }
 \end{equation}
 for the Green's function. 
 Here, we have used the abbreviations $\Jt = \frac{1}{2}J$ and
 $z_{\up,\dn} = \pm 1$. All the interesting physics 
 is included in the memory matrix 
 \begin{equation}
   \label{eq:POM-memory}
   M_{\bk\ps}(\omega) 
       = \Lbra{Q \Lio\, \cd{\bk\ps}} \frac{1}{\omega - Q \Lio Q}
         \Lket{Q \Lio\, \cd{\bk\ps}} ,
 \end{equation}
 which again has the structure of a resolvent, resulting in a 
 form of $G_{\bk\ps}$ in (\ref{eq:POM-GF}) involving continued fractions. 

 The expression for the memory matrix cannot be treated exactly. 
 However, at this stage we are only aiming at a perturbational 
 expansion of the self-energy $\Sigma_{\bk\ps}$ in the form
 \begin{equation}
   \label{eq:SOPT}
   \Sigma_{\bk\ps} = - \Jt \,z_\ps \!\CF{S^z} 
                   + \Jt^2 \,\gamma_{\bk\ps} + \ldots.
 \end{equation}
 This allows some simplifications.
 In (\ref{eq:POM-memory}) already the $\Ham_{\rm sf}$-contribution 
 in $| Q \Lio \cd{\bk\ps})$ gives rise to a factor $\Jt^2$. Hence, any
 approximation of the Liouville operator in the denominator
 is still correct in this order and is thus consistent with 
 (\ref{eq:SOPT}). A {\it conventional} SOPT 
 implies a replacement of $\Lio$ by its free part $\Lio_0$. A
 SOPT {\it relative to HF} is given by a Liouville operator that
 corresponds to the Hamiltonian
 \begin{equation}
   \label{eq:Ham2}
   \Ham_0^{\rm (MF)} = \Ham_0 - \sum_{\bk,\ps} \Jt \,z_\ps \!\CF{S^z} \n_{\bk\ps} .
 \end{equation}
 In both cases we obtain a similar result
 \begin{eqnarray}
   \label{eq:2nd_result_MF}
   \gamma_{\bk\ps}
     &=& - \CF{S^z}^2 G_{\bk\ps}^{\rm (0/MF)}
         + \frac{1}{N^2} \sum_\bq \CF{S^z_{-\bq} S^z_\bq}
           G_{\bk+\bq,\ps}^{\rm (0/MF)} \\\nonumber
     &+& \! \frac{1}{N^2} \sum_\bq \left\{
              \CF{S^\ms_{-\bq} S^\ps_\bq}
            + 2 z_\ps \CF{S^z_{\bz} \n_{\bq+\bk,\ms}}
           \right\} G_{\bk+\bq,\ms}^{\rm (0/MF)},
 \end{eqnarray}
 where \( S^\ps_\bq = S^x_\bq + \ie z_\ps S^y_\bq \).
 For the expectation values contained in (\ref{eq:2nd_result_MF}) we make 
 use of the fact that we aim for a result correct to second order in 
 $\Jt$ and evaluate them using the eigenstates of the free/mean-field system. 

 A {\it self-consistent} SOPT on the other hand can be obtained in the
 same manner as suggested by Bulk and Jelitto \cite{BuJ90} for the 
 Hubbard model. Within this procedure the unperturbed part is altered in each
 iteration cycle by the memory-matrix of the previous cycle:
 \begin{equation}
   \label{eq:sc-cycle}
   \Ham_0^{(N+1)} = \Ham_0^{(N)} + \sum_{\bk\ps} M^{(N)}_\ps \n_{\bk\ps}.
 \end{equation}
 It turns out that this procedure is equivalent to a replacement of 
 the Green's functions at
 the right side of (\ref{eq:2nd_result_MF}) by the full propagators
 as obtained in the previous iteration cycle.
 In (\ref{eq:sc-cycle}) we use the additional approximation that 
 the memory-matrix is summed over $\bk$ and hence only a local 
 self-energy is considered.

\section{An exact solution}
\label{sc:exact}

 The model can be restricted to the limit of a ferromagnetically 
 saturated semiconductor. This limit
 is characterized by two mathematical consequences: 
 Firstly, a semiconductor is defined by an empty conduction band
 at zero temperature,
 hence $\CF{\ldots \cn{\bk\ps}}=0$. Secondly, ferromagnetic
 saturation leads to trivial spin expectation values:
 \( \CF{\ldots S_\bq^z}=NS \CF{\ldots}, \CF{\ldots S_\bp^+}= 0 \).
 An application of these simplifications to the result
 (\ref{eq:2nd_result_MF}) yields the self-energy
 \begin{equation}
   \label{eq:SOPT-Polaron}
   \Sigma_{\bk\ps} = \Sigma_\ps = - \Jt \,z_\ps S 
                   + \Jt^2 \,2S \frac{1}{N} \sum\limits_\bq 
                   G_{\bq\ms}^{\rm (0/MF)} \,\delta_{\ps\dn} .
 \end{equation}

 One can see directly that in this limit a {\it self-consistency} 
 iteration does not yield any further results.
 This is because the $\ps=\dn$ Green's function is
 uniquely determined by $\ps=\up$ propagators, which have the
 iteration-independent self-energy $\Sigma_\up \equiv - \Jt S$.

 Due to the restrictions of this limit the memory matrix 
 can in fact be treated more accurately. Following the intention of
 continued-fraction expansion the memory matrix (\ref{eq:POM-memory})
 can itself be considered as a resolvent, to which 
 the concept of the POM is applied.
 \begin{equation}
   \label{eq:mem_as_res}
   M (\omega)
    = \frac{1}{\omega-\left[\hat\Omega + \hat{M} (\omega) \right]
                        \hat\chi^{-1} } \hat\chi .
 \end{equation}
 The higher-order memory matrix $\hat{M}$ will have a form
 similar to that given in (\ref{eq:POM-memory}). Again the
 Liouville operator $\Lio$ in the denominator should be
 approximated to allow for an analytical solution of the
 associated geometric series. According to {\it conventional}
 perturbation theory it is replaced by $\Lio_0$, the action
 of a free, undisturbed system of electrons. 
 After sophisticated calculations, which will be published
 elsewhere, the self-energy is obtained as
 \begin{equation}
   \label{eq:result3}
   \Sigma_{\ps} = -\Jt \,z_\ps S 
                   + \delta_{\ps\dn} \,\Jt^2 \,
                     \frac{ 2S \frac{1}{N} \sum\limits_\bq 
                   G_\bq^{(0)} (\omega)}{1 
                   - \Jt(1-S) \frac{1}{N} \sum\limits_\bq G_\bq^{(0)} (\omega) }.
 \end{equation}
 This is certainly an improvement of (\ref{eq:SOPT-Polaron}) and
 contains the previous result if expanded up to order $\Jt^2$. 

 Even though we called it perturbation theory, it is however not correct for
 the next order in $\Jt$. The exact $\Jt^3$-contribution to the
 self-energy can actually be shown to be
 \begin{equation}
   \label{eq:J3contribution} 
   \Sigma_\dn^{(3)} = \Jt^3 \, 2S \left\{
     \bigg[  \frac{1}{N} \sum\limits_\bq G_\bq^{(0)} (\omega) \bigg]^2 \!\!
     - \frac{S}{N} \sum\limits_\bq \left[G_\bq^{(0)} (\omega) \right]^2
   \right\}.
 \end{equation}
 As a matter of fact, the second sum in (\ref{eq:J3contribution}) is a 
 diverging contribution. This already becomes apparent if one looks at its
 imaginary part, rewrites the $\bk$-sum as an integral over the free
 DOS, separates a Lorentzian and considers the fact that $0^+$ is 
 infinitesimally small.  
 We were able to show that after a summation over all orders in $J$ the
 diverging terms cancel. Nevertheless, (\ref{eq:J3contribution}) 
 demonstrates that for the s-f model a strict perturbation theory is
 only possible up to second order in $J$. The different orders
 of an expansion of the continued fraction within the POM
 apparently do not have this limitation.

 The result (\ref{eq:result3}) can be further improved, if a 
 perturbation theory {\it relative to HF} is chosen. Without going 
 into the details we just provide the result as:
 \begin{equation}
     \label{eq:result3-HF}
   \Sigma_{\ps} = -\Jt \,z_\ps S 
                   + \delta_{\ps\dn} \,\Jt^2 \,
                     \frac{ 2S \frac{1}{N} \sum\limits_\bq 
                   G_{\bq\ms}^{\rm (MF)} (\omega)}{1 
                   - \Jt \frac{1}{N} \sum\limits_\bq G_{\bq\ms}^{\rm (MF)} (\omega) }.
 \end{equation}
 This expression is actually identical to the result of an exact
 calculation, where the Liouville operator has not been reduced or 
 altered. Within the POM we were able to perform the
 derivation of the memory matrix (\ref{eq:POM-memory}) using the complete
 operator $\Lio = \Lio_0 + \Lio_{\rm sf}$. However,
 it is not necessary to give the lengthy calculations here since its
 result (\ref{eq:result3-HF}) has already been verified by other methods.
 \cite{MM68,STU91,ShM81,NDM85,NMJR96}

 The self-energy (\ref{eq:result3-HF}) corresponds to an exact eigenstate of
 the Hamiltonian (\ref{eq:Ham1}). For the spin-down electrons
 this eigenstate, which is the ground state \cite{STU91} for
 antiferromagnetic coupling ($J<0$), is called magnetic polaron 
 \cite{ShM81}. Its interesting and nontrivial dynamical features, 
 which give rise to a scattering part and a quasi-particle peak in the 
 density of states have been discussed in detail by Nolting et al. \cite{NMJR96}.
 Apparently, it is possible to retrieve these features within the 
 projection-operator formalism. The reason why already an approximation
 yields the correct result is the fact that the resulting
 two-dimensional Liouville subspace is sufficient to completely 
 describe the physics of a ferromagnetically saturated semiconductor.

\begin{figure}[th]    
    \centerline{\epsfig{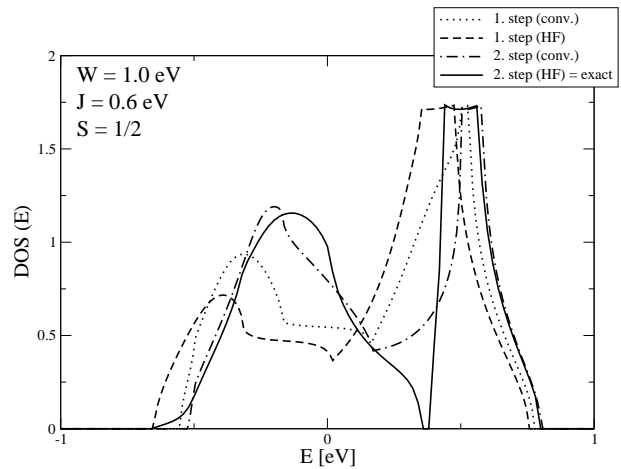}}
    \caption[Comparison of DOS for polaron]
    {Comparison of the $\dn$-QDOS for different self-energies in the
      limit $n=0$ and $\CF{S^z}=S$. 
      The dotted and dashed line are the results of SOPT
      (\ref{eq:SOPT-Polaron}), resp. {\it conventional} and {\it relative to
      HF}. The dashed-dotted and the solid line give the 
      corresponding results (\ref{eq:result3})/(\ref{eq:result3-HF})
      for the next step of the POM. 
     \label{fg:Polaron} }
 \end{figure}

 Now we have obtained several approximate forms of the electronic
 self-energy. Formula (\ref{eq:SOPT-Polaron}) provides an expression for
 the first and second order in the coupling constant $\Jt$. The
 (diverging) third order is given in 
 (\ref{eq:J3contribution}). An improvement of the SOPT is given in
 (\ref{eq:result3}) on a {\it conventional} way and in
 (\ref{eq:result3-HF}) {\it relative to HF}. One can compare these
 self-energies by looking at their quasiparticle densities of
 states (QDOS). 
 Obviously, the spin-$\up$ spectrum is always a mean-field-shifted free DOS.
 Hence, we can limit ourselves to the spin-$\dn$ spectrum, 
 which is shown in figure \ref{fg:Polaron}. 

 In this figure a relatively large $J = 2 \Jt$ has been chosen to reveal
 the differences more clearly. If one compares the DOS of the {\it
 conventional} SOPT and the SOPT {\it relative to HF} with that of the
 exact solution, one gets the impression that the first one is
 the better approximation. However, if one compares these
 two approaches for the second step of the POM (equivalent to a
 larger relevant Liouville subspace) it is clear that the version {\it
 relative to HF} has to be preferred, since only this one gives the exact
 result. Nevertheless, result (\ref{eq:result3}) is already a 
 satisfactory approximation.
 As mentioned above a {\it self-consistent} calculation is redundant for
 the discussed limit. 

\section{Generalization to arbitrary configurations}
\label{sc:arbitrary}

 After these considerations on the magnetic polaron we return
 to the discussion of the second order perturbation theory (SOPT) 
 given in formula (\ref{eq:2nd_result_MF}).
 As argued before, the limit of a ferromagnetically saturated
 semiconductor can be used to check the quality of this result. 
 In this limit the SOPT-result (\ref{eq:SOPT-Polaron}) turns out to
 be only a poor approximation of the exact solution as demonstrated in
 figure \ref{fg:Polaron}.
 Since the magnetic polaron is indeed an important feature
 of the s-f model, ways of improving (\ref{eq:SOPT},\ref{eq:2nd_result_MF})
 should be considered. 

 In the previous section we explained how an improvement can 
 be achieved within this particular limit.
 A proper application of an additional step within the 
 projection operator method finally leads to the expressions 
 (\ref{eq:result3}) and (\ref{eq:result3-HF}). We generalize 
 the analytical structure of these results to the following 
 {\it ansatz} for arbitrary band occupations:
 \begin{equation}
   \label{eq:MPT}
     \Sigma_{\bk\ps} (E) = -\Jt\,z_\ps\!\CF{S^z} 
                   + \Jt^2 \frac{ a_{\bk\ps} \gamma_{\bk\ps}(E)}
                            { 1 - b_{\bk\ps} \gamma_{\bk\ps}(E)} .
 \end{equation}
 Although obtained in a completely different manner, this is exactly 
 the kind of a modified perturbation theory \cite{KK96} (MPT)
 which has turned out to be the most promising analytical approach to
 the Hubbard model \cite{PHWN98}. 
  
 For the s-f model it has two advantages: First, it does not destroy
 the correctness of the second order term proportional to $\Jt^2$, but
 gives the freedom to fit the parameters $a_{\bk\ps}$ and
 $b_{\bk\ps}$ such that further criteria are fulfilled.
 Secondly, since the SOPT-result for the self-energy
 (\ref{eq:SOPT}) automatically reproduces
 the first three moments of the corresponding Green's function correctly,
 the choice \( a_{\bk\ps} = 1 \) will ensure the same
 for the {\it ansatz} (\ref{eq:MPT}).

 It remains to determine the parameter \( b_{\bk\ps} \). The most
 straightforward choice merely ensures the correctness of the 
 ferromagnetically saturated semiconductor ( \( b_{\bk\ps} = \Jt/2 S \) ).
 However, the resulting densities of states have unphysical
 features. Furthermore, we have learned from the Hubbard model 
 \cite{PHWN98} that a
 fit to the spectral moments of the Green's function yields more 
 promising results. A \( b_{\bk\ps} \) which is determined by
 \begin{equation}
   \label{eq:b-choice-Moments}
   b_{\bk\ps} = \Jt^2 
      \frac{ \Lbra{Q \Lio \cd{\bk\ps}} \Lio \Lket{Q \Lio \cd{\bk\ps}} 
           - \Lbra{Q \Lio \cd{\bk\ps}} \Lio_0 \Lket{Q \Lio \cd{\bk\ps}} }
           { \LSP{Q \Lio \cd{\bk\ps}}{Q \Lio \cd{\bk\ps}}^2 } 
 \end{equation}
 ensures that the third coefficient of the high-energy expansion 
 of (\ref{eq:b-choice-Moments}) is identical to the one
 of the exact self-energy. 
 As explained in more detail in appendix \ref{High-energy expansion}
 this fit is correct for the first four moments of the Green's
 function.

 In an MPT which is based on a {\it conventional} SOPT the Liouville
 operator $\Lio_0$ in (\ref{eq:b-choice-Moments}) is understood to 
 correspond to the free part of the Hamiltonian $\Ham_0$.
 Then only the $\Jt^3$-contribution in the numerator of $b_{\bk\ps}$
 remains to be evaluated and one obtains
 \begin{equation}
   \label{eq:b-Moments-ev}
   b_\ps = \Jt 
     \frac{ \left[ S(S+1) - z_\ps \CF{S^z} - \CF{S^z}^2 \right]
            ( z_\ps \CF{S^z} + 1 ) + q_\ps }
          { \left[ S(S+1) - z_\ps \CF{S^z} - \CF{S^z}^2 + 2 p_\ps
            \right]^2 },
 \end{equation}
 where $p_\ps$ and $q_\ps$ are sets of further correlation functions,
 which are given in appendix \ref{Abbreviations}, 
 but have the property to vanish in the limit $n \to 0$. 
 It is instructive to study this limit in more
 detail. On the one hand it can be combined with the additional
 constraint of ferromagnetic saturation $\CF{S^z} = S$. If the 
 obtained $b_\dn$ is placed into the MPT-{\it ansatz} (\ref{eq:MPT}), then
 the self-energy becomes identical to the one given in (\ref{eq:result3}).
 On the other hand, one can consider the zero-bandwidth situation 
 \( \eps_\bk \ \equiv T_0 \) within the limit $n \to 0$. For a
 dispersion-less Green's function all summations in expression
 (\ref{eq:2nd_result_MF}) for $\gamma_{\bk\ps}$ can readily be
 performed and in this limit the self-energy becomes
 \begin{equation}
   \label{eq:S-Moments-4}
   \Sigma_\ps = - \Jt \, z_\ps\!\CF{S^z} + \Jt^2 
   \frac{[S(S+1) - z_\ps\!\CF{S^z} - \CF{S^z}^2] \frac{1}{E-T_0} }
        {1- \Jt(z_\ps\!\CF{S^z} + 1) \frac{1}{E-T_0} }.
 \end{equation}
 Comparing this form of $\Sigma_\ps$ to the result of an exact 
 calculation available for the zero-bandwidth limit \cite{NoM84,NRRM01} 
 reveals that the expression is correct.

 For a SOPT {\it relative to HF} the expression for $\Lio_0$ in 
 equation (\ref{eq:b-choice-Moments}) contains an additional
 term according to the mean-field 
 contribution \( \Sigma_\ps = - \Jt z_\ps \CF{S^z} \) in 
 (\ref{eq:Ham2}). This yields a correction $\delta b_\ps =$
 \begin{equation}
   \label{eq:delta_b}
      \frac{ \Sigma_\ps \big[
               \CF{(S^z)^2} - \CF{S^z}^2
             \big]
           + \Sigma_\ms \big[
               \CF{S^\ms S^\ps} + 2 z_\ps \CF{S^z} n_\ms
             \big]
           }
           { \left[ 
               S(S+1) - z_\ps \CF{S^z} - \CF{S^z}^2 + 2 p_\ps
             \right]^2 
           } 
 \end{equation}
 to the former result (\ref{eq:b-Moments-ev}). We will focus again 
 on the limit $n \to 0$. If combined with the additional constraint 
 of ferromagnetic saturation, the exact result (\ref{eq:result3-HF})
 is obtained. A consideration of the zero-bandwidth limit 
 yields the following expression for the self-energy
 \begin{eqnarray}
   \label{eq:S-Moments-6}
   \Sigma_\ps &=& - \Jt \, z_\ps\!\CF{S^z} + \Jt^2 Y \\ \nonumber 
              &+& \Jt^2 
   \frac{[S(S+1) - z_\ps\!\CF{S^z} - \CF{S^z}^2] \frac{1}{E-T_0+ \Jt
       z_\ms\!\CF{S^z}} }
        {1- \Jt \left(\frac{1}{E-T_0+ \Jt z_\ms\!\CF{S^z}} + X \right) } .
 \end{eqnarray}
 Apart from the correction terms $X$ and $Y$ this is again the exact
 result (\ref{eq:S-Moments-4}).  However, both terms are
 proportional to expressions, which vanish in the paramagnetic regime. 
 Therefore, they also vanish for the zero-bandwidth limit for
 which the assumption of any finite magnetization does not lead to
 consistent results. \cite{Hic01}

 In the discussion so far we have tested our MPT-{\it ansatz} (\ref{eq:MPT})
 in the limit $n \to 0$. One can repeat the same transformations for
 the opposite case $n \to 2$. By doing this one will notice, that the
 same formulae are obtained. The only difference is the change of the
 sign of $\ps$ and of $b_{\bk\ps}$. This is due to particle-hole
 symmetry in the system. Therefore, in the same sense as for $n=0$ our 
 MPT-{\it ansatz} (\ref{eq:MPT}) fulfills the limit of the magnetic polaron
 and the zero-bandwidth limit for $n=2$.

\section{Self-Consistent Results}
\label{sc:numerics}

 Using the MPT approach of the previous section we have 
 the possibility to generalize the improvement to the POM for the 
 limit of a ferromagnetically saturated semiconductor to arbitrary 
 parameter regimes. In section \ref{sc:exact} we argued that
 a {\it self-consistent} calculation is redundant in this limit. 
 This does not hold for the generalized version. Here $\gamma_{\bk\ps}$,
 as given in (\ref{eq:2nd_result_MF}), does not vanish for 
 spin$-\up$ electrons and consists of propagators for both kinds of 
 spin-directions. Consequently, we have performed a {\it self-consistent} 
 numerical iteration of the self-energy. This was carried out along the
 lines sketched at the end of section \ref{sc:method}. Additionally,
 the MPT-parameter $b_\ps$ has to be adjusted such that 
 $\Sigma_\ps$ in (\ref{eq:delta_b}) describes the full self-energy
 and not only its mean-field part. 

 The details of this procedure are discussed in section 
 \ref{sc:correct}. If properly performed it is an 
 ``upgrading'' of the perturbation theory {\it relative to HF}
 in the sense that its properties are maintained.
 In particular,  the evaluation for the ferromagnetically
 saturated semiconductor yields the previous and exact results of the 
 magnetic polaron. Additionally, the atomic limit is fulfilled 
 for the empty and completely filled conduction band, and the 
 particle-hole symmetry of the system is conserved by the ansatz. 
 In other regimes it is ensured that all results are correct 
 at least to order $\Jt^2$. However, because of the self-consistency 
 the method incorporates more correlations and scattering 
 effects than a straightforward second-order perturbation theory 
 description does.
 For all these reasons we believe that the {\it self-consistent} 
 MPT is not only correct in the weak-coupling regime, but also 
 for moderate values of $\Jt$. 

\begin{figure}[t]    
    \centerline{\epsfig{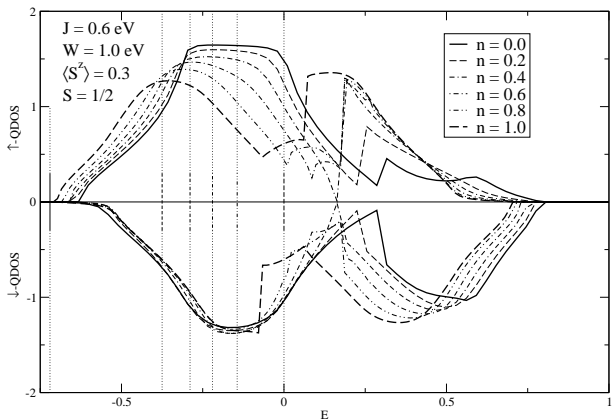}}
    \caption[Dependence on band occupation for sc MPT]
    {Dependence on band occupation $n$ for a {\it self-consistent}
      MPT. For free electrons a sc-DOS with bandwidth $W$ is chosen. 
      The parameters are as given. The vertical lines indicate
      the positions of the Fermi energy, to allow for the different
      values of the band occupation. 
      \label{fg:n-dependence} }
 \end{figure}

\begin{figure}[t]    
    \centerline{\epsfig{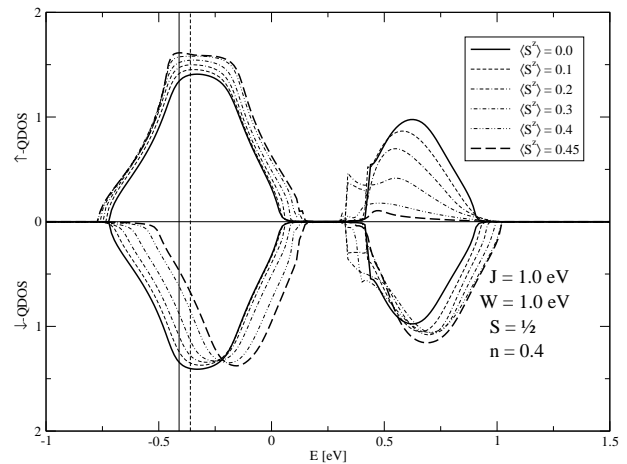}}
    \caption[Dependence on magnetization for sc MPT]
    {Change of the QDOS with the magnetization $\CF{S^z}$
     for a {\it self-consistent} MPT.  
     The position of the Fermi energy for the highest and lowest
     $\CF{S^z}$ is marked by vertical lines.  
     \label{fg:T-dependence} }
 \end{figure}

 We discuss the results in terms of the quasiparticle densities of
 states (QDOS). According to our {\it ansatz} the QDOS is correct 
 for $n=0, \CF{S^z}=S$, however its variation with a change
 of these parameters is of particular interest. Figure 
 \ref{fg:n-dependence} shows the dependence on the first parameter. 
 The dependence on the magnetization (connected to 
 temperature via a Brillouin function) is given in figure \ref{fg:T-dependence}. 
 In both cases a medium value has been chosen for the fixed parameter, 
 respectively. 
 It goes without saying that our calculations are also thermodynamically
 self-consistent. The iteration ensures that the values 
 of the correlation functions are consistent with the obtained 
 one-particle Green's function. Additionally, the chemical potential 
 is adjusted to the desired particle number. Its position 
 is indicated by vertical lines in the figures.

 In particular the change of $n$ in figure \ref{fg:n-dependence} 
 has remarkable consequences for the QDOS. The clear dependence
 on the filling of the conduction band points out strong 
 correlation effects, induced by the coupling $\Jt$. 
 For $n=0$ the structure of the QDOS is closely related to the 
 ferromagnetically saturated semiconductor. In particular the 
 spin-$\up$ spectrum has the shape of the free (simple cubic) DOS
 and the scattering part of the spin-$\dn$ spectrum can 
 clearly be seen. Only the polaron subband shows a deformation,
 due to finite-lifetime effects. Excited spin-$\up$ electrons can 
 enter the energy region of the polaron, flip their spin and
 absorb a magnon since we are not close to saturation.

 If the chemical potential (and accordingly the band occupation)
 is increased, the spectral weight is 
 redistributed between both subbands. For the chosen set of parameters 
 the changes with $n$ are most noticable in the spin-$\up$ QDOS, 
 where the upper subband steadily increases in importance at 
 the expense of the lower subband. A sharp jump in the QDOS 
 close to the pseudo-gap remains a striking feature for all values 
 of $n$. It is also interesting to note, that the lower band 
 edge is shifted by some $0.1$ eV in the spin-$\up$ QDOS, 
 whereas it remains at almost the same position for the
 spin-$\dn$ QDOS. This behavior is very much different 
 in an MPT {\it relative to HF}\cite{HSN03} and is a hint that 
 in the {\it self-consistent} MPT it are mainly the majority-spin electrons 
 that experience strong correlations. 

 As the band occupation approaches half filling ($n=1$), the 
 point-symmetric form of the QDOS nicely represents the particle-hole 
 symmetry of the system. The character of the upper
 spin-$\dn$ subband becomes identical to the lower spin-$\up$ subband,
 since the latter is the polaron band for $n=2$. 
 For the same reason we skip the  plots for $n>1$, they can 
 be obtained from the band occupations $2-n$.

 A higher value for the coupling strength $\Jt$ is chosen
 in figure \ref{fg:T-dependence}. Therefore, the scattering and the 
 polaron subband are well separated already for a nearly 
 saturated system. The gap remains present for all temperatures
 employed in the calculations. There are only small changes of the position 
 of the bands as a function of the magnetization. 
 Nevertheless, the edge of the lower spin-$\up$ subband
 shifts to lower energies if the temperatures 
 is lowered from $T=T_C$ ($\CF{S^z}=0$) to smaller values $T \to 0$ 
 (maximum $\CF{S^z}$). For semiconductors such an effect is known as 
 the red shift of the optical absorption edge. In metals, since
 the lower spin-$\dn$ subband is shifted in the opposite direction, 
 it leads to a polarization of the conduction electrons of over 60\%. 
 The existence of energy regions well below the Fermi edge 
 occupied entirely by majority-spin electrons 
 is a remarkable result. Similar effects have also been reported in 
 other approximations\cite{IK94} when studying half-metallic 
 ferromagnets. However, the continuous shift of the chemical 
 potential with magnetization prevents 100\% polarization of the 
 conduction electrons in our calculations.
 The dependence on the chemical potential is such that
 the effect disappears completely for smaller values ($n \to 0$), 
 where the $\up$-QDOS and the $\dn$-QDOS occupy the same energy 
 region (see figure \ref{fg:n-dependence}).

 The situation at the lower edge of the upper subband in figure 
 \ref{fg:T-dependence} is 
 less systematic. An extra peak in the spin-$\up$ QDOS obtains
 its maximum for $\CF{S^z} \approx 0.3$ and then vanishes again. The physical 
 interpretation of this feature is not yet clear. In this context
 we also have to mention that the choice of the parameter $b_\ps$ 
 with its explicit dependence on the self-energy in (\ref{eq:delta_b}) 
 substantially increases the numerical effort. In the energy region just 
 mentioned it is particularly difficult to obtain convergence. 

 The distribution of the spectral weight, on the other hand, 
 is indisputable. It shows the transition from the ferromagnetically
 saturated configuration (dashed lines), which even for $n=0.4$ clearly 
 displays the features of the exact solution in section \ref{sc:exact}, 
 to the paramagnetic regime (solid lines), which has to be 
 symmetric with respect
 to the $x$-axis. Again the more profound changes are observed for
 the majority-spin electrons. The increasing spectral weight of 
 the upper spin-$\up$ subband can be explained with higher
 magnon-numbers in this regime. 

 An artifact of our method is the fact that for both 
 spin directions two subbands are always obtained. With other 
 approaches\cite{NRR03}, one sometimes observes a third band.
 This is explained by atomic-limit calculations, where
 for finite band-occupations always three out of four subbands 
 have non-vanishing spectral weight. It needs further modifications of 
 our method to retain these features. At the present stage the atomic 
 limit is only correct for $n=0$ and $n=2$. 

\section{The proper method}
\label{sc:correct}

 The last point brings us to an assessment 
 of our {\it self-consistent} approach. Apart from the above 
 mentioned catalogue of analytical properties a 
 comparison with other conceivable weak-coupling approaches
 is desirable. 

 A comparison with an MPT which uses $\gamma_{\bk\ps}$ 
 obtained by {\it conventional} SOPT 
 is straightforward. From the analytical considerations 
 in section \ref{sc:arbitrary} we can conclude that 
 even an MPT based on an SOPT {\it relative to HF} should be 
 preferred as compared to one based on a {\it conventional} SOPT. 
 This is because the former correctly
 incorporates the important limiting case of the ferromagnetically 
 saturated semiconductor, whereas the latter does only 
 reproduce, in this limit, the less accurate expression 
 (\ref{eq:result3}). The discussion of the atomic limit does not
 provide an argument in favour of one the approaches, 
 since at the end both yield physical expressions of the same quality. 
 In these limits our {\it self-consistent} approach has 
 the same properties as the MPT {\it relative to HF}. 
 
 \begin{figure}[th]    
    \centerline{\epsfig{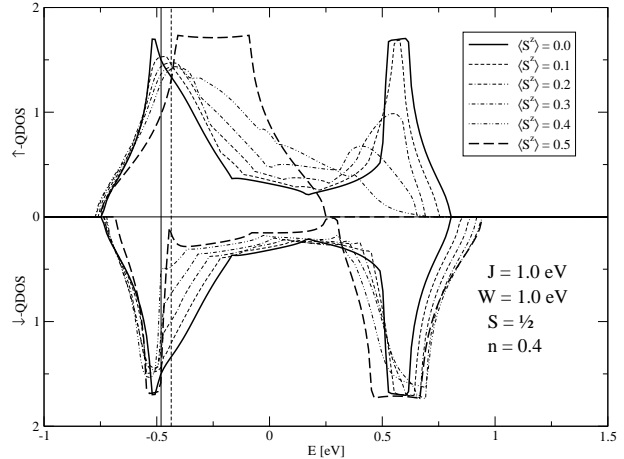}}
    \caption[Comparison of DOS for MPT-HF]
    {Change of the QDOS with the magnetization $\CF{S^z}$. The
     MPT-calculations are based on a SOPT {\it relative to
     HF}. The parameters are chosen as in figure \ref{fg:T-dependence}.
     \label{fg:MPT-HF} }
 \end{figure}
 Our comparison with an MPT {\it relative to HF} is based on 
 numerical results with this method. For the same parameters
 as in figure \ref{fg:T-dependence} we obtain the set of QDOS 
 given in figure \ref{fg:MPT-HF}. 
 Even though its main features look sound again, there are a 
 set of minor aspects which make this approach questionable. 
 The most obvious drawback are prominent and unexplainable peaks 
 close to the Fermi energy. The functional dependence is
 not smooth and dominated by the free DOS. Also for magnetizations below
 saturation a gap in the spin-down QDOS is expected from other 
 theories\cite{MSN01,NRRM01},  but this does not exist in this approach. 
 Additionally it is noteworthy that the onset of the spin-up QDOS starts
 for smaller energies as compared to that for the spin-down QDOS. 

 In the {\it self-consistent} MPT of the previous section most of 
 these peculiarities
 are not present. A comparison with figure \ref{fg:T-dependence}
 permits the conclusion that the iteration of the self-energy 
 yields shapes of the QDOS which are broader and smoother. 
 This can be understood analytically and is closely related 
 to the fact that the self-energy becomes complex by iteration. 
 These findings make the {\it self-consistent} version more reliable.

\begin{figure}[t]    
    \centerline{\epsfig{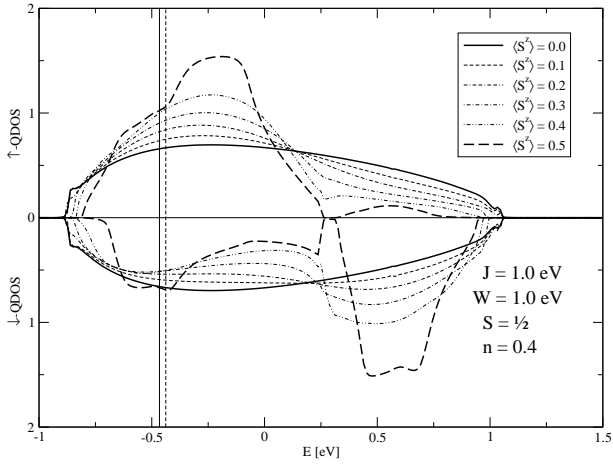}}
    \caption[Comparison of DOS for sc SOPT]
    {Dependence on magnetization $\CF{S^z}$ for an MPT based
     on a {\it self-consistent} SOPT. The parameters are as in figure 
     \ref{fg:T-dependence}.
      \label{fg:MPT-V2} }
 \end{figure}

 However, there are several possibilities to incorporate 
 self-consistency into the MPT. An ambiguous point is the 
 order of the applied steps. In contrast to the calculations of the
 previous section 
 one could start with the expression of the MPT as obtained in 
 an approach {\it relative to HF} and continue by dressing all 
 included propagators as full Green's functions. Hence, the
 parameters $b_{\bk\ps}$ are fixed, which is still correct 
 for a fit to the high-energy expansion up to order $E^{-2}$. 
 The consequences for the densities of 
 states are shown in figure \ref{fg:MPT-V2}.
 They look reasonable for values of $\CF{S^z}$ close to saturation. 
 However, the single, broad, elliptic band which emerges close
 the paramagnetic regime is a surprising feature. Not only is this result
 inconsistent with other approximations of the s-f model 
 \cite{NRRM01,NRMJ97,HV00,NRR03}, moreover it is incorrect in
 the zero-bandwidth limit. This is because the same structure 
 remains in the limit $W \to 0 / n \to 0$, where two narrow bands
 around $-\Jt S$ and $\Jt (S+1)$ are expected as indicated by exact
 calculations\cite{NoM84}. The shortcoming can already be seen 
 analytically when looking at equations (\ref{eq:S-Moments-4})
 or (\ref{eq:S-Moments-6}). If the propagators are dressed without
 a change of $b_\ps$, then an equality is not possible with the 
 exact solution:
 \begin{equation}
   \label{eq:AL-check1}
   \Jt^2 \, \frac{S(S+1)\frac{1}{E-T_0 -\Sigma }}
              {1- \Jt \frac{1}{E-T_0 -\Sigma }}
   \neq
   \Jt^2 \, \frac{S(S+1)\frac{1}{E-T_0} }
              {1- \Jt \frac{1}{E-T_0} } = \Sigma .
 \end{equation}

 This conflict can be resolved if the order of arguments is
 changed. Now the starting point is a self-consistent SOPT and 
 only afterwards the result is fitted to the high-energy expansion. 
 The consequence is not a higher accuracy in powers of $E^{-1}$ but
 an additionally dressed fitting parameter $b_\ps$. According
 to the correction given in (\ref{eq:delta_b}) it now contains 
 the full self-energy and not only its mean-field contribution.
 This leads to the correct result for the zero-bandwidth limit
 at $n=0$:
 \begin{equation}
   \label{eq:AL-check2}
   \Jt^2 \frac{S(S+1)\frac{1}{E-T_0 -\Sigma }}
              {1- (\Jt-\Sigma) \frac{1}{E-T_0 -\Sigma }}
   =
   \Jt^2 \frac{S(S+1)\frac{1}{E-T_0} }
              {1- \Jt \frac{1}{E-T_0} } .
 \end{equation}

 \begin{figure}[t]    
    \centerline{\epsfig{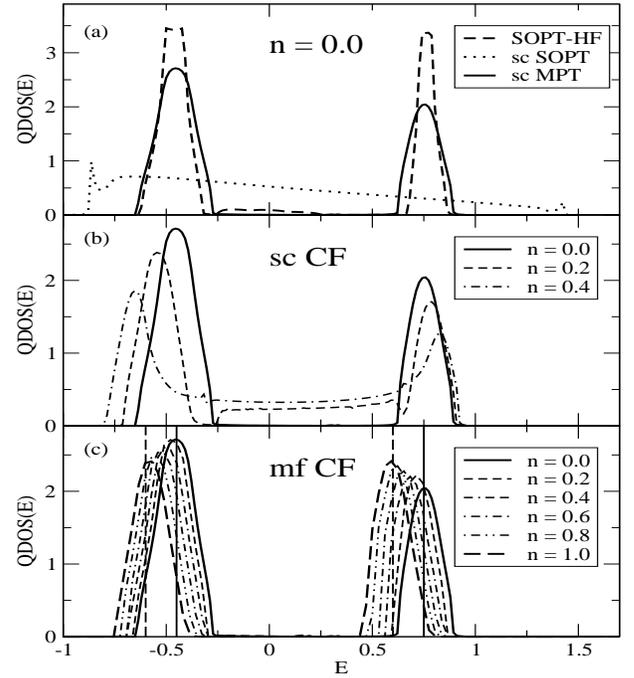}}
    \caption[Discussion of atomic limit]
    {Comparison of different MPT approaches for parameters
     close to the atomic limit (bandwidth $W=0.5$ eV, $J=0.6$ eV, 
     $S=3/2$, $\CF{S^z}=0.0$). {\bf (a)} dashed line: the MPT is 
     based on a SOPT {\it relative to HF}, dotted line: only the SOPT
     input is treated {\it self-consistently} and the parameter $b_\ps$ 
     is not altered, solid line: the {\it self-consistency} has also 
     consequences for $b_\ps$. For the latter there are two
     versions to treat the correlation functions (CF) in $b_\ps$: either
     self-consistent {\bf (b)} or mean-field like {\bf (c)}. 
     The vertical lines indicate the energy-positions of the
     maximums of the subbands.
     \label{fg:MPT-AL} }
 \end{figure}
 Additionally, it has serious consequences for all other parameter
 regions. We compare numerical results for the three different
 approaches mentioned above in figure \ref{fg:MPT-AL}a. The set of
 parameters $W, \Jt, S$ is chosen to be close to the atomic limit. 
 The choice $n=0, \CF{S^z}=0$ ensures exactness in this limit. 
 If the input $\gamma_{\bk\ps}$ to the MPT (\ref{eq:MPT})
 is the result of the SOPT {\it relative to HF} (dashed line) then 
 two nearly free subbands at the correct positions are obtained.
 There seems to be a third flat band between them. 
 To dress only the propagators and not the fitting parameter $b_\ps$
 (dotted line) is definitely wrong. However, one can 
 convince oneself, that an inclusion of the full self-energy in
 the calculation of $b_\ps$ (solid line) really yields a considerable 
 improvement for $n=0$. Here one again observes the two narrow
 subbands, the excitation energies for an electron that aligns
 its spin parallel ($-\Jt S$) or antiparallel ($\Jt(S+1)$) to the
 localized spin.

 Nevertheless, there remains an uncertainty in the determination 
 of $b_\ps$ as far as the correlation function $p_\ps$
 in equation (\ref{eq:b-Moments-ev}) and (\ref{eq:delta_b}) is
 concerned.  Its definition
 is chosen such that it can be calculated with the help of
 the one-particle Green's function $G_{\bk\ps}$:
 \begin{equation}
   \label{eq:p_ps}
   p_\ms = - \frac{1}{\pi \hbar \Jt N} \sum_\bk \int\limits_{-\infty}^\infty
           (E - \eps_\bk) \cdot 
           \frac{ \Im{\rm m} G_{\bk\ps} }{ {\rm e}^{\beta E} + 1}
           \,{\rm d} E.
 \end{equation}
 This form of the spectral theorem has the handicap to be only applicable 
 for determining a sum of correlation functions (\ref{df:p_ps}). Whenever 
 being confronted with a single correlation function contribuation to 
 (\ref{eq:p_ps}), we are forced to perform a mean-field
 decoupling:   \( \CF{S^z n_\ms} = \CF{S^z}\cdot\CF{\n_\ms} \).
 This is still compatible with the expansion in $\Jt$ and $E^{-1}$. 
 It is therefore straightforward to treat those correlation functions
 for which relations such as (\ref{eq:p_ps}) exist as accurate as possible,
 and perform approximations for the remaining correlation functions.

 The consequences of this methodology for the atomic limit are shown 
 in figure \ref{fg:MPT-AL}b. The QDOS looks sound for $n=0.0$, but 
 shows a broad non-quasiparticle structure between the subbands
 for $n>0$. Its spectral weight increases with band occupation
 at the latter's expense. This dependence is qualitatively different
 compared to that of a third, intermediate band in the QDOS for the SOPT 
 {\it relative to HF},
 which remains small for all values of $n$. Here, already for 
 a band occupation of $n=0.4$ the gap is completely filled. 
 It is difficult to find a physical explanation for such a behavior. 
 Small satellite peaks between the subbands were also reported
 for other approximation methods\cite{NRRM01} and were attributed to
 un-trapped electrons which experience the global magnetization 
 $\CF{S^z}$ as an effective quantization axis. A shift of the
 spectral weight within the intermediate structure as a function of
 the net magnetization has also been observed in our calculations.
 However, the missing symmetry in the paramagnetic regime and 
 the strong dependence on the band occupation does not fit into 
 this picture. The accompanying shift of the two subbands is also
 surprising. For these reasons we believe that these features
 are an artifact of the approximations used as we know
 that the atomic limit is only correctly included for $n=0$. 

 In contrast to the methodology to determine $b_\ps$ as accurately 
 as possible, it has apparently a much higher priority to treat 
 all included correlation functions on an equal footing. Figure
 \ref{fg:MPT-AL}c shows results for the QDOS with the same set
 of parameters as in figure \ref{fg:MPT-AL}b. The only modification
 in the theory is a mean-field decoupling of all correlation 
 functions in $b_\ps$. The effects on the QDOS is dramatic, 
 as the intermediate structure completely vanishes now. Additionally,
 the shift of the two subbands happens in a comprehensible way: 
 Due to the particle-hole symmetry we expect for $n=2$ two subbands 
 at positions $-\Jt (S+1)$ and $+\Jt S$. Since our approach apparently 
 only allows for a single
 bandgap, the change from an $n=0$ to an $n=2$ configuration 
 can only be implemented by the system if the two peaks move
 continously to their new positions. Accordingly their positions 
 at half filling ($n=1$) have to be
 \( E_{\pm} = \pm \Jt (2S+1)/2 \), as seen in the figure.  
 Also the redistribution of the spectral weight takes place along 
 these lines. 

 Based on our experience with the MPT we draw the following conclusion:
 The most promising weak-coupling approach to the periodic s-d(f) exchange 
 model is a {\it self-consistent} MPT. Expression 
 (\ref{eq:2nd_result_MF}) dressed with full propagators $G_{\bk\ps}$
 should be used as the input from second-order perturbation 
 theory. Only afterwards the parameters $a_{\bk\ps}$ and 
 $b_{\bk\ps}$ in the MPT-ansatz (\ref{eq:MPT})
 should be determined such that the high-energy expansion is 
 fulfilled to power $E^{-2}$, which implies that $b_\ps$ carries 
 a dependence on the full self-energy. For the correlation functions
 entering $b_\ps$ it is important that they are all treated on the
 same footing. This is, at this stage, only possible by using a 
 mean-field decoupling.

\section{Summary and Outlook}
\label{sc:summary}

 Within the presented work we have demonstrated how the projection
 operator method can be exploited to find an analytical approach
 to the periodic s-d(f) exchange model and that it is indeed a valuable tool 
 in this context. Nevertheless, we argued that the second 
 order perturbation is insufficient and suggested an improvement 
 in the form of an MPT. The principle structure of this ansatz 
 results from a study of the limit of the ferromagnetically 
 saturated semiconductor. We showed that the calculations have
 to be performed {\it self-consistently}. In a further step
 a clarification of the proper treatment of the fitting parameter
 $b_\ps$ was necessary. At the end we were able to make an 
 informed statement, which of all possible approaches is
 the most reliable one. On the one hand it is satisfactory that a 
 certain approach was able to produce considerably better results
 than other attempts. On the other hand, the high sensitivity
 of the QDOS to the methodology used to treat the
 correlation functions includes the danger of arbitrariness. 

 All of the suggested approaches have in common that they are correct up 
 to second order in the coupling parameter $J$. Nevertheless, our
 improvements are non-perturbative in the sense that no Taylor expansion
 is provided. On the one hand this is for technical reasons, since the
 absence of a Wick's theorem for spin operators significantly
 complicates the calculation of Feynman diagrams. On the other 
 hand already Shastry and
 Mattis \cite{ShM81} argued that a perturbation theory in $J$ would
 fail because of the discontinuities in the physical properties 
 of the model as $J$ changes sign. Additionally, we have pointed out 
 that the $J^3$-contribution to the exact self-energy in the
 limit of a ferromagnetically saturated semiconductor diverges although
 the sum over all orders yields a finite result. 

 The qualitative properties of the densities of states presented here
 are very similar to the findings of other approaches\cite{NRMJ97}. 
 Since the former results were based on decoupling schemes for
 Green's functions the approximations incorporated into these
 calculations are difficult to control in their quality. 
 With our results we can confirm {\it a posteriori}
 and justify these findings. This includes a complete 
 set of strong correlation effects discussed there.

 However, in its present state the documented method is only
 an approach to the electronic part of the s-f model.
 Whenever correlation functions that carry a dependence on the
 properties of localized magnetic moments emerged we had to 
 perform some crude approximations. It is connected to this fact, that
 we only considered a $\bk$-independent self-energy by taking 
 the average over the whole Brillouin zone. 
 In this direction their is certainly room for further improvements. 

\section*{Acknowledgments}
 One of us (T.~H.) gratefully acknowledges the financial support of the 
 {\it Friedrich-Naumann Foundation}. This work also benefitted from the
 support of the {\it SFB 290} of the {\it Deutsche Forschungsgemeinschaft}.

\begin{appendix}
 \section{High-energy expansion}
 \label{High-energy expansion}

 The high-energy expansion of the Green's function can
 be obtained from its representation as a resolvent 
 (\ref{eq:POM-GF}):
 \begin{equation}
   \label{eq:HEE-GF}
   G_{\bk\ps} = \sum_{l=0}^\infty \frac{m_{\bk\ps}^{(l)}}{\omega^{l+1}}
              = \sum_{l=0}^\infty \frac{
                   \Lbra{\cd{\bk\ps}} \Lio^l \Lket{\cd{\bk\ps}}
                }{\omega^{l+1}}
 \end{equation}
 The coefficients $m_{\bk\ps}^{(l)}$ are called spectral moments and are
 determined by an $l$-fold commutator with the Hamiltonian $\Ham$. 
 Apart from the mean-field contribution to the frequency matrix
 (\ref{eq:POM-frequency}), the self-energy is identical to the
 memory matrix (\ref{eq:POM-memory}). Hence, its high-energy 
 expansion 
 \( \Sigma_{\bk\ps} = \sum\limits_{m=0} C_{\bk\ps}^{(m)}/\omega^m \)
 is given by
 \begin{equation}
   \label{eq:HEE-SE}
   \Sigma_{\bk\ps} = -\Jt z_\ps\!\CF{S^z} 
   + \sum_{m=0}^{\infty} \frac{\Lbra{Q \Lio \cd{\bk\ps}}  
         (Q\Lio Q)^m  \Lket{Q \Lio \cd{\bk\ps}}}{\omega^{m+1}} .
 \end{equation}
 Due to the properties of $Q$, the coefficients $C_{\bk\ps}^{(m)}$ can 
 be expressed in terms of spectral moments:
 \begin{eqnarray}
   \label{eq:HEE_C1}
   C_{\bk\ps}^{(1)} &=& m_{\bk\ps}^{(2)} - \left[ m_{\bk\ps}^{(1)} \right]^2 \\
   \label{eq:HEE_C2}
   C_{\bk\ps}^{(2)} &=& m_{\bk\ps}^{(3)} - 2 m_{\bk\ps}^{(2)}
                        m_{\bk\ps}^{(1)} + \left[ m_{\bk\ps}^{(1)}
                        \right]^3 .
 \end{eqnarray}
 However, in our second-order perturbation theory the Liouville 
 operator $\Lio$ is replaced by a part $\Lio_0$:
 \begin{eqnarray}
   \label{eq:HEE-SOPT}
   \Sigma_{\bk\ps}^{\rm (SOPT)} &\!\!\!&= -\Jt z_\ps\!\CF{S^z} 
         + \Jt^2 \gamma_{\bk\ps} \\ \nonumber
   =&\!\!\!& -\Jt z_\ps\!\CF{S^z} 
   + \sum_{m=0}^{\infty} \frac{\Lbra{Q \Lio \cd{\bk\ps}}  
         (Q\Lio_0 Q)^m  \Lket{Q \Lio \cd{\bk\ps}}}{\omega^{m+1}} .
 \end{eqnarray}
 If this result is used for the MPT (\ref{eq:MPT}), the high-energy 
 expansion of the self-energy is given by:
 \begin{equation}
   \label{eq:HEE-SE-MPT}
   \Sigma_{\bk\ps}^{\rm (MPT)} = -\Jt z_\ps\!\CF{S^z} 
      + \Jt^2 a_{\bk\ps} \gamma_{\bk\ps} 
      + \Jt^2 a_{\bk\ps} b_{\bk\ps} \left[ \gamma_{\bk\ps} \right]^2 
      + \ldots
 \end{equation}
 A comparison with the exact expression (\ref{eq:HEE-SE}) shows,
 that $\gamma_{\bk\ps}$ is correct to order $\omega^{-1}$. To 
 ensure correctness to the same order for $\Sigma_{\bk\ps}^{\rm (MPT)}$
 the parameter $a_{\bk\ps}$ has to be chosen as $1$. To order
 $\omega^{-2}$ the self-energy $\Sigma_{\bk\ps}^{\rm (MPT)}$
 has the coefficient
 \begin{equation}
   \label{eq:HEE-2nd-MPT}
   \Lbra{Q \Lio \cd{\bk\ps}} \Lio_0 \Lket{Q \Lio \cd{\bk\ps}}
   + \frac{1}{\Jt^2} b_{\bk\ps}
     \LSP{Q \Lio \cd{\bk\ps}}{Q \Lio \cd{\bk\ps}}^2 .
 \end{equation}
 In order to ensure that also this coefficient is exact, it has to be
 equal to 
 \begin{equation}
   \label{eq:HEE-2nd-SE}
   C_{\bk\ps}^{(2)} = \Lbra{Q \Lio \cd{\bk\ps}} \Lio \Lket{Q \Lio \cd{\bk\ps}}
 \end{equation}
 Equality can be obtained if $b_{\bk\ps}$ is chosen as 
 suggested in equation (\ref{eq:b-choice-Moments}) above. As can
 be seen from (\ref{eq:HEE_C2}) in the high-energy this order 
 expansion implies the correctness of the four moments 
 \( m_{\bk\ps}^{(0)}, \ldots, m_{\bk\ps}^{(3)} \) of the Green's
 function.

 \section{Abbreviations}
 \label{Abbreviations}

 For the sake of brevity we have introduced some shorthand notations
 in this papers. The full expressions are given here. 

 Equation (\ref{eq:b-Moments-ev}):
 \begin{eqnarray}
   \label{df:p_ps}
   p_\ps &=& z_\ps \CF{S^z n_\ms} - \CF{S^\ms \cd{\ps} \cn{\ms} } \\
   \label{df:q_ps}
   q_\ps &=& 4  z_\ps \CF{S^z} p_\ps -  2 \left( x_\ps - p_\ps \right)
   \\
   \label{df:x_ps}
   x_\ps &=& S(S+1) \CF{\n_\ps} - z_\ps \CF{S^z \n_\ps}
       + 2 z_\ps \CF{S^z \n_\ps \n_\ms} \nonumber\\&&
       + \CF{ (S^z)^2(\n_\ms - \n_\ps)} 
       - z_\ps \CF{S^\ps S^z \cd{\ms} \cn{\ps} + \mbox{h.c.}} \nonumber\\&&
       - \CF{S^\ps \cd{\ms} \cn{\ps} + \mbox{h.c.}}
 \end{eqnarray}
 We have evaluated $x_\ps$ by making use of the equivalence
 \begin{eqnarray}
   \label{eq:xrelation}
     \sum_j T_{lj} \left( \CF{S^\ps_l \cd{l\ms} \cn{j\ps}}
                        - \CF{S^\ps_l \cd{j\ms} \cn{l\ps}}
                   \right) \nonumber\\
   = -\Jt \left(x_\ps - S(S+1) \CF{n_\ms} - p_\ps \right)
 \end{eqnarray}
 and arguing that the left hand side vanishes for almost all
 parameter configurations and in particular if mean-field
 decoupling is applied. 

 Equation (\ref{eq:result3}):
 \begin{eqnarray}
   \label{df:X}
   X &=& \Jt Z \, \frac{ \CF{(S^z)^2} - \CF{S^z}^2 }
                    {S(S+1) \!- z_\ps\!\CF{S^z} - \CF{S^z}^2} \nonumber\\
     &-& 2 \Jt \, \frac{ z_\ps \CF{S^z}\CF{(S^z)^2} - z_\ps\CF{S^z}^3 }
                    {S(S+1) \!- z_\ps\!\CF{S^z} - \CF{S^z}^2}     \cdot
                \frac{1}{E-T_0-\Jt z_\ps\!\CF{S^z}} \nonumber\\
     &+& 2 \Jt Z z_\ps\!\CF{S^z} \left[ \frac{ \CF{(S^z)^2} - \CF{S^z}^2 }
                  {S(S+1) - z_\ps\!\CF{S^z} - \CF{S^z}^2} \right]^2 \\
   Y &=& Z \, \frac{ \CF{(S^z)^2} - \CF{S^z}^2 }
                { 1- \Jt \left[ \frac{1}{E-T_0-\Jt z_\ps\!\CF{S^z}} + X
                         \right] } \\
   Z &=& \frac{1}{E-T_0+\Jt z_\ps\!\CF{S^z}}
        -\frac{1}{E-T_0+\Jt z_\ms\!\CF{S^z}} 
 \end{eqnarray}
 $\CF{S^z}$ vanishes in the paramagnetic regime. Hence, $Z$ becomes zero,
 which implies the same for $X=Y=0$. 

\end{appendix}

%\bibliography{/home/hickel/Publications/References/KLM,/home/hickel/Publications/References/SOPT}

\end{document}